\documentclass[preprint,showpacs,preprintnumbers,amsmath,amssymb,nofootinbib]{revtex4}

\usepackage{etex}
\usepackage{graphicx}
\usepackage{dcolumn}
\usepackage{bm}
\usepackage{amssymb,amsthm,amscd,amsbsy,array}
\usepackage{amsmath,dsfont}
\usepackage{soul} 
\usepackage{graphics,xcolor}

\numberwithin{equation}{section}

\newcommand{\half}{{\scriptstyle{\frac{1}{2}}}}
\def\2{{\frac{1}{2}}}
\newcommand{\const}{\mathop{\rm const}\nolimits}

\def\bR{{\mathds{R}}}
\def\bbeta{{\bm{\beta}}}

\def\tchi{{\widetilde{\chi}}}
\def\bgamma{{\bm{\gamma}}}

\def\bomega{{\bm{\omega}}}

\def\bp{{\bm{p}}}
\def\chio{{\chi^{\mathrm{o}}
}}
\def\tbx{\widetilde{\bm{x}}}
\def\ba{{\bm{a}}}

\def\br{{\bm{r}}}

\def\bE{{\bm{E}}}
\def\bB{{\bm{B}}}

\def\bp{{\bm{p}}}

\def\bS{{\bm{S}}}
\def\bs{{\bm{s}}}

\def\hbp{{\widehat{\bm{p}}}}

\def\bx{{\bm{x}}}

\def\by{{\bm{y}}}

\def\beq{\begin{equation}}
\def\eeq{\end{equation}}
\def\beqa{\begin{eqnarray}}
\def\eeqa{\end{eqnarray}}
\def\nn{\nonumber}
\def\barray{\left(\begin{array}}
\def\earray{\end{array}\right)}
\def\barraynb{\begin{array}}
\def\earraynb{\end{array}}

\def\SO{{\rm SO}}
\def\ort{{\mathfrak{o}}}

\def\smallover#1/#2{\hbox{$\textstyle\frac{#1}{#2}$}} %

\newcommand{\bbR}{\mathbb{R}}

\newcommand{\cE}{{\mathcal{E}}}
\newcommand{\Tr}{\mathrm{Tr}}

\newcommand{\belle}{\boldsymbol{\ell}}

\newcommand{\cR}{\mathcal{R}}

\usepackage{color}
\newcommand{\red}{\textcolor{red}}  
\newcommand{\blue}{\textcolor{blue}}

\begin{document}

\preprint{arXiv:1508.02188v5 [hep-th] 
}

\title{Helicity of spin-extended chiral particles
}

\author{
M. Elbistan$^{1}$\footnote{mailto:elbistan@impcas.ac.cn.},
C. Duval$^{2}$\footnote{mailto:duval@cpt.univ-mrs.fr},
P. A. Horv\'athy$^{1,3}$\footnote{mailto:horvathy@lmpt.univ-tours.fr},
P.-M. Zhang$^{1}$\footnote{e-mail:zhpm@impcas.ac.cn},
}

\affiliation{
${}^1$ Institute of Modern Physics, Chinese Academy of Sciences, Lanzhou, (China) 
\\
${}^2$
Aix-Marseille University, CNRS UMR-7332,  Univ. Sud Toulon-Var
13288 Marseille Cedex 9,
(France)
\\
${}^3$ Laboratoire de Math\'ematiques et de Physique
Th\'eorique,
Universit\'e de Tours,  (France)
}

\date{\today}

\begin{abstract}
The helicity of a free  massless relativistic particle, $\chio={\bf s}\cdot{\bf p}/|{\bf p}|$, is generalized, for a particle in an electromagnetic field,
 to $\chi={\bf s}\cdot{\bf p}/\cE$ where $\cE$ is the modified kinetic energy. Both $\chio$ and $\chi$ coincide and are conserved for minimal coupling (gyromagnetic ratio $g=0$) but are different and neither of them is conserved when the coupling is non-minimal, $g\neq0$, generating non-zero effective mass. For a chiral particle with $g=2$ in a constant electric field both helicities converge asymptotically to the same value. Helicity is also conserved for minimal gravitational coupling.
\\[30pt]
\noindent
Phys.\ Lett.\ {\bf A 380} 1677 (2016)
\end{abstract}


\maketitle

\tableofcontents

\section{Introduction}

The spin of free photons can only be parallel or anti-parallel to their momentum \cite{Feynman}. Accordingly, their \emph{helicity}, defined as the eigenvalue of the spin operator projected onto the momentum  \cite{HELI,Tung}, can only be $\pm \hbar$.

Recently, much attention has been devoted to \emph{chiral fermions}, which are  massless relativistic particles with spin $\pm 1/2$ \cite{StephanovYin,SonYama2,Stone,QunWang,Manuel,DHchiral}.
 An underlying semiclassical model (we call the \emph{c-model}) has been deduced from the Weyl equation \cite{StephanovYin} and used to simplify complicated quantum field theoretical calculations. 
In the {c-model} spin is ``enslaved'' to the momentum, i.e., the spin term 
 is $\half\hbp$. The
 twisted Lorentz symmetry of the {c-model} \cite{ChenSon} could be explained by embedding into Souriau's massless spinning particle \cite{SSD,DHchiral,DEHZ-I}, our  \emph{S-model}. The latter has two additional degrees of freedom represented by an ``unchained'' spin vector, $\bs$, whose \emph{projection onto the momentum} satisfies, in the free case,
\beq
\chio=\bs\cdot \hbp  = \pm|{s}|\,. 
\label{oldhelicity}
\eeq
Classically $s$, the scalar spin  \cite{SSD,DHchiral}, can be any real number, but quantization requires  $s= n/2,\, n=0, \pm1,\pm2,\dots$ in units of $\hbar$ \cite{SSD,HPAix}. Moreover, (geometric) quantization allows one to recover the Weyl equation for $|s|=1/2$ \cite{DuvalQuant,SSD}, and the Maxwell equations for $|s|=1$ \cite{SSD}, respectively.
$\chio$  is  the \emph{classical helicity} \cite{SSD,DHchiral}.

The trajectory of a free massless particle in the spin-enlarged S-model is ill-defined, as  the free motions  are 3-dimensional surfaces, acted upon by a three-parameter group called  ``Wigner-Souriau (WS) translations" \cite{HELI,SSD,DHchiral,DEHZ-I,Stone3,Stone:2015kla}.
The latter can also be used to ``enslave'' the spin to the momentum, $ \bs \to s\, \hbp$,   allowing us to establish the equivalence of the free c and S - models \cite{DHchiral,DEHZ-I}. 
 
Chiral fermions can also be coupled to an electromagnetic field, which changes the situation dramatically~: the coupled S and c models behave \emph{substantially differently} \cite{DHchiral,EM-HP-g0}. 
In this Letter we study the S-model, for which we \emph{postulate} that the spin --  field coupling should endow the initially massless system with an \emph{effective mass} (square), 
\beq
P_{\mu}P^{\mu}=-({eg}/{2})S. F\,,
\label{effmass}
\eeq
where $F=(F_{\mu\nu})$ is electromagnetic tensor, $S.F= S^{\mu\nu}F_{\mu\nu}$ is the spin-field coupling, and $g$ is the gyromagnetic factor \cite{Duval3these,DFS,Sour74}.
WS translations  get broken and the 3D motion-surfaces condensate into world lines. Spin enslavement becomes inconsistent with the coupled S-dynamics and the  spin $\bs$ is unchained (whereas, in the c-model, spin is still enslaved) \cite{DHchiral,EM-HP-g0}. This issue will not be discussed henceforth.
In this Letter we point out 
 that when the effective mass does not vanish, 
\emph{let alone the helicity constraint} (\ref{oldhelicity}) can be maintained.

\section{The massless spinning model 
}

Souriau's relativistic classical massless spinning model
(the S-model) is described by a $9$-dimensional evolution space $V^9$ labeled with Minkowski 4-vectors $R=(R^\mu)$ (position in space-time), $P=(P^\mu)$ (four-momentum) and the spin tensor $S=(S_{\mu\nu})\in \ort(3,1)$, subject to the constraints \cite{Duval3these,Kun,DFS,Sour74,DHchiral,AnAn}\footnote{The mass term could be an {arbitrary} function of $S.F$ \cite{Duval3these,Kun,DFS, Sour74,AnAn} ; the particular form (\ref{effmass}) is suggested by its validity for the Dirac equation, which has $g=2$.}  
\beq
S_{\mu\nu}P^{\nu}= 0\,,
\qquad
\half{S}_{\mu\nu}S^{\mu\nu}=s^2\,,
\qquad
P_{\mu}P^{\mu}=-({eg}/{2})\,S. F\,.
\label{constraints}
\eeq

The rather complicated  equations of motion, 
\# (5.17)  in \cite{DHchiral}, which are the zero rest mass analogs of the Bargmann-Michel-Telegdi equations for  massive relativistic particles \cite{BMT}, are valid in full generality including non-constant fields. They are given by kernel of the closed two-form
\begin{equation}
\sigma
=
-dP_\mu\wedge{}dR^\mu
-\smallover{1}/{2s^2}\,d{S}^\mu_\lambda\wedge{S}^\lambda_\rho\,d{S}^\rho_\mu
+\half{}eF_{\mu\nu}\,dR^\mu\wedge{}dR^\nu.
\label{gsigma}
\end{equation} 	
In a Lorentz frame 
$
R=(\br,t),\,
P=(\bp,\cE),\,  
S_{ij}=\epsilon_{ijk}s_k,\, S_{i4}=\big(\displaystyle\frac{\bp}{\cE}\times\bs\big)_i
$ 
where $\bs=(s_i$) is the spin vector. We stress that
the dispersion relation 
\beq
\cE=\sqrt{|\bp|^2-(eg/2)S\cdot F} =
\sqrt{|\bp|^2-2(eg/2) \bs\cdot\big(\bB-\frac{\bp}{\cE}\times\bE\big)}\;
\label{SFcE}
\eeq 
is \emph{not} an Ansatz, but follows from the posited mass formula (\ref{effmass}) and can not be changed  at will therefore \footnote{A spin-field correction term to the energy, which is in fact the weak-field approximation of our (\ref{SFcE}), was proposed in \cite{Manuel,ChenSon}, who argue that it is indeed necessary for Lorentz covariance.}.

For the sake of simplicity, hence we limit ourselves to constant external fields, $F=(\bE,\bB)$. 

For $g=0$ (minimal coupling) the above-mentioned general eqns  simplify to
\begin{subequations}
\begin{align}
(\hbp\cdot\bB)\frac{d\br}{dt}&=\bB-\hat{\bp}\times\bE,
\label{Seqsa}
\\
(\hbp\cdot\bB)\frac{d\bp}{dt}&=e(\bE\cdot\bB)\hat{\bp},
\label{Seqsb}
\\
(\hbp\cdot\bB)\frac{d\bs}{dt}&=\bp\times\bB-\bp\times(\hat{\bp}\times \bE).
\label{Seqsc}
\end{align}
\label{g0eqns}
\end{subequations}  
This case  has been analyzed in \cite{DHchiral,DEHZ-I,EM-HP-g0}: the  helicity constraint, (\ref{oldhelicity}),  holds true, but spin enslavement is broken.

For the \emph{non-minimal} value $g=2$, to which our investigations will be mostly focused, the equations of motion simplify again, namely to 
\begin{subequations}
\begin{align} 
 \displaystyle\frac{d\br}{dt}&=\displaystyle\frac{\bp}{\cE}\,,
\label{ourreq} 
\\[8pt]
\displaystyle\frac{d\bp}{dt}&=e\left(\bE+\displaystyle\frac{\bp}{\cE}\times\bB\right),
\label{ourpeq}
\\[8pt]
\displaystyle\frac{d\bs}{dt}&=\displaystyle\frac{e}{\cE}
\left(\big(\displaystyle\frac{\bp}{\cE}\times\bs\big)\times\bE
+\displaystyle{\bs}\times\bB\right).
\label{ourseq}
\end{align}
\label{cfg2}
\end{subequations}
Unlike as in (\ref{Seqsa})  which is ``purely anomalous'', no anomalous velocity term arise for $g=2$~\footnote{The anomalous terms in the general equations \# (5.17) of \cite{DHchiral} have indeed a coefficient $(g-2)$; see also~\cite{AnAn} in the planar case. 
The anomalous velocity term is actually missing from the original BMT equations~\cite{BMT}, corrected in \cite{SSD}, eqn. \# (15.35) and analyzed in detail in \cite{Duval3these} and in \cite{Sour74}, Sec. 9--10.
The possibility of having such a term was raised in \cite{AdamsBlount}.}.
Equations (\ref{cfg2}), which are reminiscent of those of a massive relativistic particle, imply that the  \emph{spin-field term is a constant of the motion} 
 \footnote{
The statement holds for any value of $g$ provided the fields are constant \cite{Sour74}.}, 
and yields also the time-variation of $\cE$,
\beq
\frac{\,d}{dt}(S.F)=0,
\qquad
\frac{\,d}{dt}(\cE^2)=2e\,{\bp\cdot\bE}.
\label{cEvar}
\eeq
Therefore the conserved energy is
$
\cE-e\br\cdot\bE\,,
$
allowing us to interpret $\cE$ in (\ref{SFcE}) as the kinetic energy.
Then, using the equations of motion (\ref{cfg2}) we find, for $g=2$,
\beq
\frac{d\chio}{dt\;\;}=
e^2\big(\frac{S.F
}{\cE^2|\bp|}\big)\,\bs\cdot\big(\hat{\bp}\times (\hat{\bp}\times\bE) \big),
\label{Ohelicityloss}
\eeq
implying that $\chio$  (\ref{oldhelicity}) is \emph{not conserved} in general. In Sec. \ref{motionSect}
below we present an explicit example, where the helicity is indeed \emph{not a constant of the motion}.

Some authors \cite{Bliokh,BMPLA,BEPL} advocate using, instead of our ``true position'' $\br$,  the ``(Newton-Wigner-)Pryce" coordinate \cite{NWP}
\beq
\tbx=\br+\frac{\hbp\times\bs}{|\bp|}\,,
\label{NWP}
\eeq
they identify with the ``center of the particle''.
In the free case $\tbx$ is  conserved and can be used to label a  motion, see   eqn. (3.19) of \cite{DHchiral}. The components the Pryce coordinate   do not commute (as it is well-known), resulting in the non-localizability of the particle. 
Using (\ref{NWP}), spin becomes ``enslaved" to the momentum (as in the c-model),
 i.e., the angular momentum is 
$ 
\belle=\tbx\times\bp+s\,\hbp\,.
$ 
Moreover, the 
 velocity relation (\ref{ourreq}) \emph{acquires anomalous terms} even for $g=2$, 
\begin{eqnarray} 
\frac{d\tbx}{dt}&=&
\frac{\bp}{\cE}
-e\frac{\bs\times\bE}{|\bp|^2}
-e(\hbp\cdot\bE)\big(\frac{2\cE^2-|\bp|^2}{\cE^2|\bp|}\big)\frac{\hbp\times\bs}{|\bp|}
+\frac{e}{\cE}
(\frac{\hbp\times\bs}{|\bp|})\times\bB.\qquad
\label{Prycevelocity}
\end{eqnarray}
The anomalous $\bE$-terms are, in particular, responsible for topological spin transport in the spin Hall effect; they are the 3D analogs of the anomalous velocity term in the plane \cite{DHexo,AnAn,ZhH,DHHMS}. 
 It is also worth stressing that further anomalous terms  arise for non-constant fields [not considered by other authors], see eqn. \# (5.17) of \cite{DHchiral}.

\section{Motion in a pure electric field for $g=2$}\label{motionSect}

In the c-context, an analytic solution could be found in a pure electric field \cite{ZhH}, and now we show that this also happens for the spin-extended model. Restricting our attention at $g=2$ throughout this section, we find that, for $\bB=0$ eqns. (\ref{cfg2})  reduce to
\beq
\cE\, 
 \displaystyle\frac{d\br}{dt}=\displaystyle{\bp},
\qquad
\displaystyle\frac{d\bp}{dt}=e\bE,
\qquad
\displaystyle\frac{d\bs}{dt}=\displaystyle\frac{e}{\cE^2}
\big(\displaystyle{\bp}\times\bs\big)\times\bE,
\label{pureE}
\eeq
with $\cE$ obtained by putting $g=2$ and $\bB=0$ into  (\ref{SFcE}).

The momentum equation is integrated as $\bp(t)=\bp_0+e\bE{t}$ and then from (\ref{cEvar}) we infer that
$ 
\cE^2(t)=\cE_0^2+2e(\bp_0\cdot\bE)t+e^2\bE^2t^2
$.  
(Thus for large $t$, we have $\cE\sim e|E|t$, consistently with the conservation of $S.F$ and 
with $|\bp| \sim e|E|t$). 
The $\bE$-component of spin is a constant of the motion,
$
{d(\bs\cdot\bE)}/{dt}=0.
$
Combining our results, 
\beq
\frac{d\bs}{dt}=-\frac{e}{\cE^2(t)}
\Big((\bs_0\cdot\bE)(\bp_0+e\bE t)-(\bp_0\cdot\bE+e\bE^2t)\bs \Big). 
\label{ESpineq}
\eeq 
Integrating this explicitly time-dependent first order differential equation yields
\beq
\bs(t)=\frac{\bs_0\cdot\bE}{\bE^2\cE_0^2-(\bp_0\cdot\bE)^2}\Big(
\cE_0^2\bE+
e(\bp_0\cdot\bE){\bE}t-(\bp(t)\cdot\bE)\bp_0\Big)+\cE(t)\,\ba,
\label{stsol}
\eeq
where the constant vector $\ba$ is determined by the initial conditions $\bp(0)=\bp_0,\,\bs(0)=\bs_0$ and $\br(0)=\br_0$; eqn. (\ref{stsol}) implies that
$\ba$ is perpendicular to the electric field. 
The  spatial motion can be dealt with similarly, 
\beq
\br(t)=\br_0+\frac{\big(\bE\times(\bp\times\bE)\big)}{e|\bE|^3}\log\left|\Big\{\frac{\bp\cdot\bE+\cE|\bE|}{\bp_0\cdot\bE+\cE_0|\bE|}\Big\}\right|+\frac{\bE(\cE-\cE_0)}{e\bE^2}\,.
\label{rtZ}
\eeq

Eqns (\ref{pureE}) can  also be studied numerically.
$\cE(t)$
satisfies a 3rd-order algebraic equation that can be solved  for each value of $\bp(t)$ and thus for each  $t$ using Cardano's formula; then the remaining equations can be integrated numerically. 

\begin{figure}[h]
\begin{center}
\includegraphics[scale=.55]{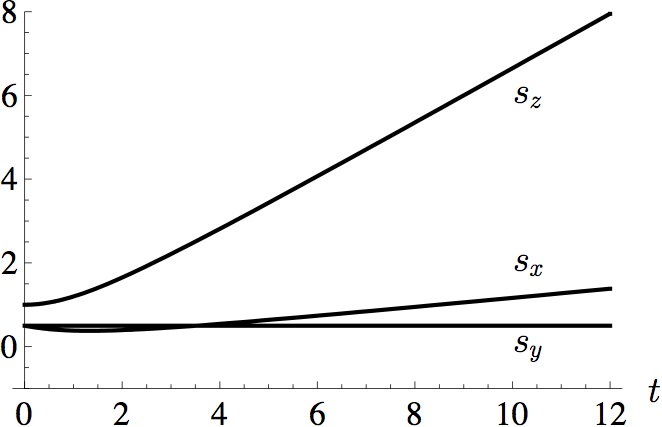}
\;\;\;\;
\includegraphics[scale=.55]{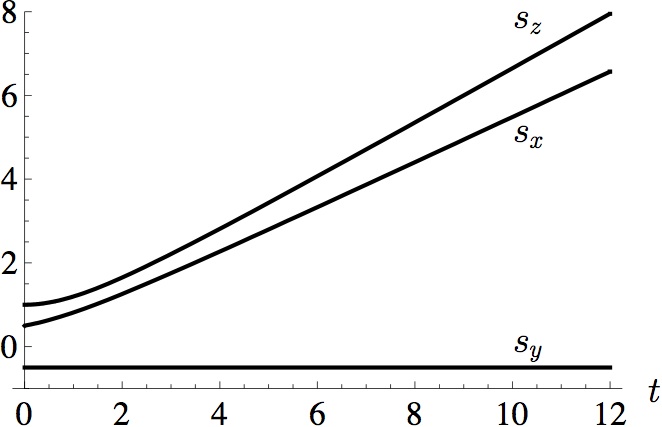}
\\
 (a) \hskip60mm (b)
\vspace{-8mm}
\end{center}
\caption{ {\it The time evolutions of the spin components, for $g=2$, 
for (not enslaved) initial conditions} 
(a)  $\bs(0)=(1/2,1/2,1)$ {\it and} (b)  $\bs(0)=(1/2,-1/2,1)$.
} 
\label{sevolution}
\end{figure}

In detail, we consider 
$\bE=E\hat{\by}$ with initial conditions $\br(0)=0,\,\bp(0)=\bp_0=p_0\hat{\bx},\, \bs(0)=\bs_0$. 
Then $\bp(t)$ stays in the $x-y$ plane and the spin evolves in the  $s_y=\const$ plane which is parallel to the $x-z$ plane and is thus \emph{perpendicular} to the  plane where the momentum moves.
Choosing
$
e=1,\ E=1,\ p_{0}=1, 
$
yields the numerical solutions plotted in Fig. \ref{sevolution}.

\begin{figure}[ht]
\begin{center}
\includegraphics[scale=.6]{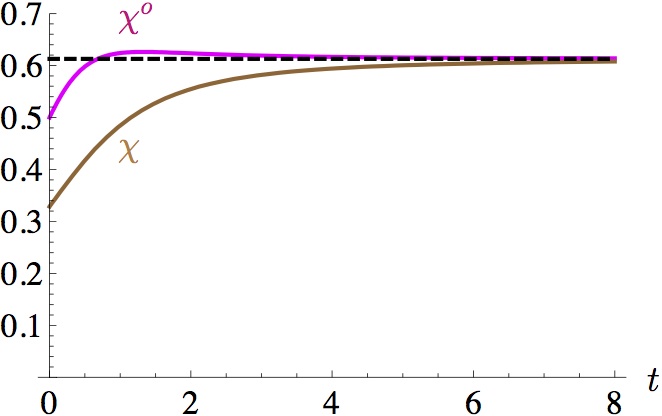}
\qquad\quad
\includegraphics[scale=.6]{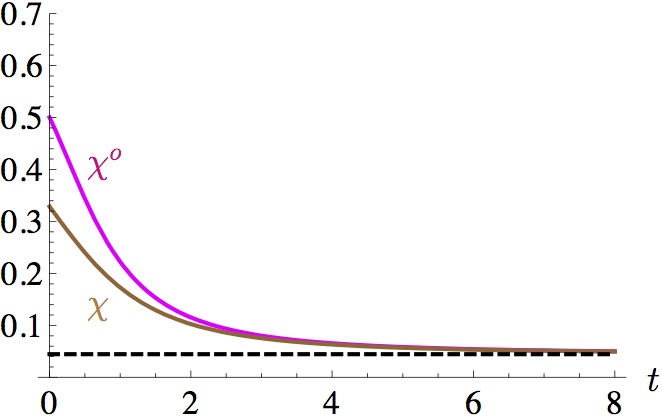}
\vspace{-2mm}
\\
\hskip-10mm (a)\hskip80mm (b)\\
\vspace{-8mm}
\end{center}
\caption{\it For a spin-extended massless particle with $g=2$ put into an electric field,
neither of the helicities ${\chio}=\bs\cdot\hbp$ nor ${\chi}=\bs\cdot\bp/\cE$ are conserved.  
However, both expressions converge to a common value asymptotically.
The  figures correspond to the same initial conditions as in Fig. \ref{sevolution}.
} 
\label{chiev}
\end{figure}

\begin{figure}[ht]
\begin{center}
\includegraphics[scale=.5]{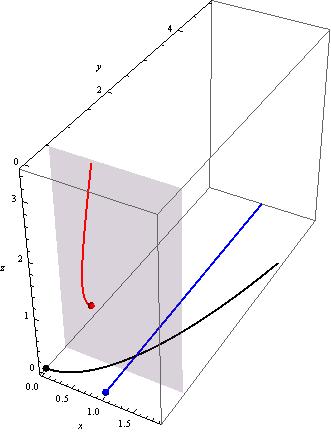}
\quad
\includegraphics[scale=.5]{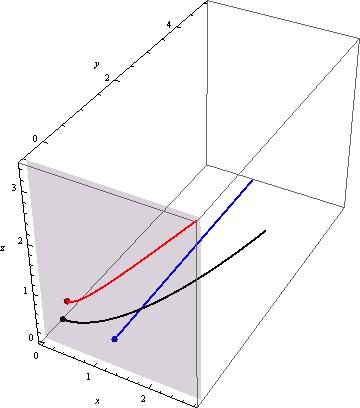}
\\
{}\hskip-10mm
 {(a)}\hskip70mm {(b)}\\
\vspace{-8mm}
\end{center}
\caption{\it 
$3$-dimensional motion for $g=2$ with our previously considered helicity non-conserving initial
conditions. \red{\bf Spin}  moves in the vertical plane $s_y=\pm\half$ whereas  \blue{\bf momentum} and {\bf position} move in  horizontal planes.
 } 
\label{rplot}
\end{figure}

Our solution above provides us with an example where  the \emph{helicity}  $\chio$  in (\ref{oldhelicity}) is \emph{not conserved}, 
as shown in Fig.\ref{chiev} by the {pink} line. 
The trajectories are depicted in Fig.\ref{rplot}.

\begin{figure}[h]
\begin{center}
\includegraphics[scale=.6]{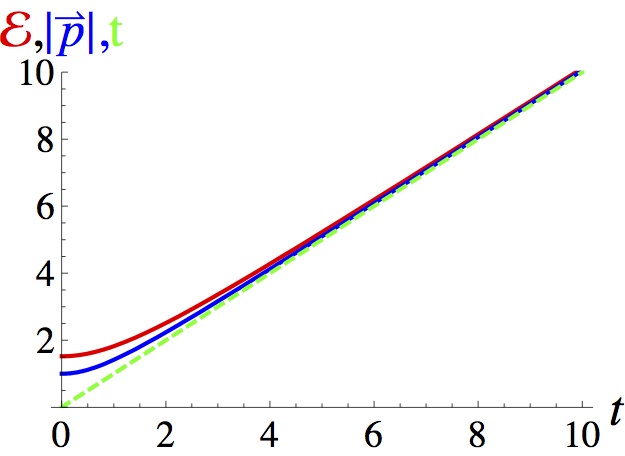}
\\
\vspace{-8mm}
\end{center}
\caption{\it Time evolution of the kinetic energy for initial conditions (b), i.e., $\bs(0)=(1/2,-1/2,1),\,\bp(0)=(1,0,0)$.
$\cE\sim|\bp|\sim {t}$ asymptotically.} 
\label{dispersion}
\end{figure}

It is worth noting that the kinetic energy, $\cE$, becomes equal to $|\bp|\sim t$ (only) asymptotically, see Fig. \ref{dispersion}.

\section{Helicity, revisited}\label{heliSc}

In this Section the gyromagnetic factor $g$ can  be any real number (unless the contrary is said). First we note that the constraint $S_{\mu\nu}S^{\mu\nu}=2s^2$  in (\ref{constraints}) implies that
\beq
\bs^2-\frac{1}{\cE^2}\big(|\bp|^2|\bs|^2-(\bp\cdot\bs)^2 \big)=s^2\,.
\label{s2}
\eeq
For $g=0$ the general dispersion relation (\ref{SFcE}) reduces to
 $\cE=|\bp|$, implying the helicity constraint $\hat{\bp}\cdot\bs=\pm{|s|} $ in (\ref{oldhelicity}). 
However for $g\neq0$ this is not so,  prompting us to revisit and generalize the  definition of helicity. 

We start again with the free case. Remember that
the space of motions  is a coadjoint orbit of the Poincar\'e group labeled by its  Casimir invariants $m$ and $s$, the mass and spin, respectively \cite{SSD}.
Decomposing the Lorentz and translational moments, $M=(M^{\mu\nu})\in\ort(3,1)$ resp. $P=(P^\mu)\in\bR^{3,1}$, into orbital and spin constituents allows us to write, 
\begin{equation}
M^{\mu\nu}=R^\mu P^\nu-R^\nu P^\mu+S^{\mu\nu}.
\label{muPoincare}
\end{equation}
In a Lorentz frame the  Casimirs are expressed as 
$ 
m^2=-\bp^2+\cE^2
$
{and}
$
{s=\belle\cdot\hbp}\,,
$ 
where $\belle$, the angular momentum, is the space component of $M$,
$M_{ij}=\epsilon_{ijk}\ell_k$ ; the real number $s$ is the scalar spin.  
The \emph{Pauli-Lubanski vector} is defined as, 
\beq
W_\sigma=\half\,\epsilon_{\mu\nu\rho\sigma}M^{\mu\nu}P^{\rho}.
\label{PLW}
\eeq
For the non-tachyonic coadjoint orbits, we have
$m\geq0$ and 
\beq
W_{\mu}W^{\mu}=-s^2m^2.
\label{Wsquare}
\eeq 
The orbits we are interested in are massless, $
m=0,
$
and have scalar spin  $|s|>0$. 
  Therefore  the null and orthogonal (thus parallel) vectors $W$ and~$P$ are proportional, 
$
W=\chi\, P, 
$
 which \emph{defines the helicity as the proportionality factor} between $W^\mu$ and $P^\mu$.  
In a Lorentz frame we find,
\beq 
W=
\chi\, 
\left(\begin{array}{c}
\bp\\
\cE
\end{array} 
\right),
\qquad
\chi=\bs\cdot\frac{\bp}{\cE}\;.
\label{Whel}
\eeq
In the free case  $\cE=|\bp|$ and we recover the old expression  in (\ref{oldhelicity}), $\chi=\chio=\pm |s|$.  Its conservation, which can also be checked using the (free) equations of motion, is consistent with the group theory, for which it is a Casimir invariant. 

Turning to the coupled case, (\ref{constraints}),
 a similar calculation yields, however
\beq
W  =
\bs\cdot\frac{\bp}{\cE}\,\left(\begin{array}{c}
\bp\\
\cE
\end{array}
\right)
-
\frac{(eg/2)S.F}{\cE}
\left(\begin{array}{c}
\bs\\
0
\end{array}\right),
\label{Wter}
\eeq
so that {Pauli-Lubanski and momentum} are no longer parallel \emph{unless the effective mass vanishes}. 
Our new helicity, ${\chi}$ in (\ref{Whel}), appears here as the coefficient of the first term. Is it conserved~? 
Rewriting (\ref{s2}) as
$ 
{\chi}^2=s^2
 +(eg/2)\frac{S.F}{\cE^2}\,|\bs|^2\,
$ 
shows that (\ref{Whel}) \emph{is conserved for minimal coupling}, $g=0$, but \emph{not} for $g\neq0$, because the extra term  is not a constant of the motion. 
For  $g=2$ a lengthy calculation shows, in particular, that (\ref{Ohelicityloss}) should be replaced by
\beq
\label{Nhelicityloss}
\frac{d{\chi}}{ dt \; }
=
-\left(\frac{e^2\,S.F}{\cE^3}\right)\,\bs\cdot\bE\,,
\eeq
which does not  vanish in general. Remarkably,
\beq
W_\mu{W}^{\mu}=s^2\,\frac{eg}{2}\,S.F\,,
\label{W2}
\eeq 
cf. the massive case above, eqn. (\ref{Wsquare})~\footnote{We just mention that one could also consider  the projection of spin onto the ``Pryce velocity'' (\ref{Prycevelocity}),
$
{\tchi}=\bs\cdot{d\tbx}/{dt}, 
$
which, for $g=0$, would be the same (and thus conserved) as $\chi=\chio$. However for $g=2$ and for $\bB=0$ we would get ${\tchi}=\chi$ which would therefore not be conserved.}.

The situation is illustrated (in {marron}) in Fig. \ref{chiev} .
Our plots indicate that both helicities ${\chio}$ and ${\chi}$ tend to the same constant value when $t\to\infty$. This is understood by  observing that, for large $t$, both the momentum and the velocity are oriented along the electric field so that $\bs\cdot\hbp \approx 
s_y(0)+\const.$ as $t\to\infty$~: helicity is not conserved  only during the short initial period.

\section{Poincar\'e invariance}

 The natural action  of the Poincar\'e Lie algebra
on space-time,
$ \delta\br=\bomega\times\br+\bbeta{}t+\bgamma,\,
\delta t=\bbeta\cdot\br+\varepsilon$ (where $\bomega,\bbeta,\bgamma$ are  infinitesimal rotations, boosts, space-translations and $\varepsilon$ a time translation)
is completed, when an electromagnetic field $F=(F_{\mu\nu})$ is switched on, by acting on the $4$-momentum and spin as 
\begin{subequations}
\begin{align}
\delta \bp &= \bomega \times \bp +  \cE\,\bbeta,
\\
\delta \cE &= \bbeta \cdot \bp,
\\
\delta \bs &= \bomega \times \bs + \bbeta \times \left(\bs \times \frac{\bp}\cE\right),
\end{align}
\label{FPaction}
\end{subequations}
respectively.
Integrating 
 (\ref{FPaction}) would yield a finite action of Poincar\'e on the evolution space $V^9$,  
$
(R,P,S) \mapsto (LR+C,LP,LSL^{-1})
$
where the Lorentz transformation $L\in\SO_+(3,1)$  and the translation $C\in\bR^{3,1}$ are the generated group elements.
 $(L,C)$  acts on the electro\-magnetic field (assumed constant) according to $F \mapsto LFL^{-1}$ since $F$  transforms as a spacetime tensor, identified here with an element of $\ort(3,1)$. It follows that the spin-field term, $S.F$, is Poincar\'e-invariant\footnote{The Poincar\'e action (\ref{FPaction}) on $V^9$ does \emph{not} preserve the two-form $\sigma$, see below. 
 },
$ 
S.F \mapsto (LSL^{-1}L).(FL^{-1}) = S.F.
$ 
Then turning to our two kinds of helicities, we find, 
\beq
\delta \chio 
= 
-\frac{(eg/2)\,S.F}{|\bp|\cE}\,(\bbeta \times \hbp) \cdot (\bs \times \hbp),
\qquad
\delta {\chi} 
= 
-\frac{(eg/2)\,S.F}{\cE^2}\, \bbeta \cdot \bs,
\label{helPoincare}
\eeq
where  $\bbeta$ is (as before) an infinitesimal Lorentz boost. Thus  \emph{neither $\chio$ nor ${\chi}$ is Poincar\'e-invariant in general}. They would,
of course, be the same [and invariant] if the effective mass $(eg/2)\,S.F$ vanished -- consistently with the frame-dependence of helicity in the massive case~\cite{Peskin}. 

\section{Conservation laws in a constant electromagnetic field}\label{NoetherSection}

Let us now work out the symmetries of a constant electromagnetic field $F$ and their associated
conserved (Noetherian) quantities. We use, once again, Souriau's framework \cite{SSD}, as outlined in 
\cite{DHchiral}. 
As already mentioned, the Souriau two-form $\sigma$  
(\ref{gsigma})  is not fully Poincar\'e-invariant. Indeed if $\sigma'=(L,C)^*\sigma$ denotes the transformed two-form and $F'=LFL^{-1}$ as before, then 
$\sigma'=\sigma+e(F'-F)$. Thus $\sigma'=\sigma$ if $F'=F$, that is,
if $LF=FL$. Denoting by $\Lambda$ an infinitesimal Lorentz transformation, the latter condition translates as $\Lambda F=F\Lambda$. Therefore an infinitesimal Lorentz transformation is symmetry of the spin-extended system iff it is a combination of the field tensor $F=(F_{\mu\nu})$ and its dual $\star(F)=(\half\epsilon_{\mu\nu\alpha\beta}F^{\alpha\beta})$,
\beq
\Lambda=\alpha F + \beta \star(F)
\label{FFdual}
\eeq
with $\alpha,\beta\in\bbR$. The resulting Poincar\'e subgroup is thus $6$-dimensional: it is the semi-direct product of the Abelian group generated by $F$ and $\star(F)$ with space-time translations.
Then the (symplectic) Noether theorem \cite{SSD}  provides us, for any value of the gyromagnetic ratio, $g$, with $6$~conserved quantities. The linear momentum
\newcommand{\pf}{\mathrm{Pf}}
\beq
\label{Pi}
\Pi^\mu = P^\mu + eF^{\mu}_{\;\nu}R^\nu
\eeq
reminiscent of  ``magnetic translations'' in the massive Landau problem  is supplemented by
 two quantities generated by  [$e$-times] $F$ and $\star(F)$ respectively, viz.,
\begin{subequations}
\begin{align}
\label{ell}
\ell&=eP_\mu{}F^\mu_{\;\nu}{}R^\nu-\half{}eF_{\mu\nu}S^{\mu\nu}+\half{}e^2\,(F_{\mu\nu}R^\nu)(F^\mu_{\;\nu}R^\nu),
\\[6pt]
\label{ellstar}
\ell^*&=eP_\mu\star(F)^\mu_{\,\nu}R^\nu-\half{}e\star(F)_{\mu\nu}S^{\mu\nu}
-\smallover1/8{}e^2\,({\star}F.F)\,R_\mu{}R^\mu.
\end{align}
\end{subequations}
In the expression of $\ell^*$ we recognize here the \emph{Pfaffian invariant} 
$ 
\pf(F)=
-\frac{1}{4}\,{\star}F.F=-\frac{1}{8}\epsilon_{\alpha\beta\rho\sigma}F^{\rho\sigma}F^{\alpha\beta}=-\bE\cdot\bB.
$ 
 In view of (\ref{Pi})  Eq. (\ref{ell}) can advantageously be rewritten as
\beq
\ell = \frac{\Pi^2}{2} +\frac{g-2}{4}\, e\,S.F, 
\eeq
highlighting the special r\^ole of $g-2$ and the fact that $S.F$ is indeed a constant of the motion.
We did not find such a nice expression for $\ell^*$ in (\ref{ellstar}).

We mention  that for $g=0$ (minimal coupling) a rather mysterious  full Lorentz symmetry arises, generalizing the (non less mysterious) rotational symmetry found in  \cite{EM-HP-g0}.

\section{Comparison with other approaches}\label{compar}

Semiclassical models can be  constructed in various ways \cite{StephanovYin,SonYama2,Stone,QunWang,Manuel,DHchiral,SSD,BMPLA,BEPL}. 
One of them is to build a model \emph{inductively}, by following some First Principles of classical mechanics.  
This has been our approach in \cite{DHchiral}, where we used Souriau's recipe which has its origin in the  theory of group representations, and starts with  postulating a symmetry  \cite{SSD}. Applied to the  massless Poincar\'e symmetry, this provides us with the \emph{free S-model} \cite{DHchiral}, whose quantization yields the Weyl equations \cite{SSD,DuvalQuant}.
 
Another frequently followed approach starts at the opposite end, i.e., \emph{deduces} an underlying classical model from some  field equations by a series of  approximations. 
In the massless case, e.g., one can start with the Weyl equation \cite{StephanovYin} and end up with the \emph{c-model},
which has no independent spin dynamics.
 Therefore the problem of helicity conservation can be addressed in the S, but not in the c context. 

Both  models can be coupled to an external gauge field. In the S-case, this is achieved by yet another First Principle, namely our effective mass formula (\ref{effmass}). 
In the deductive approach one uses instead    minimal coupling $p_\alpha\to p_\alpha-eA_\alpha$ before deriving the underlying (semi)classical c-model from the field theory.

The above-mentioned deductive approach has actually been considered earlier in the massive case, starting with the Dirac equation \cite{BMPLA} and keeping first-order terms in $\hbar$. 
In the  massless limit $m\to0$ the c-model is obtained; the purely electric solution 
  of \cite{BMPLA}  matches the one found in \cite{ZhH} in the c-context.

Further extension of the approach yields a set of rather complicated-looking equations, \# (14) and (19) in \cite{BEPL}. Putting $c=1$,  $m=0$, $E_p\approx |\bp|$ and separating the terms of order $\hbar$, the latter can be rewritten as \footnote{
Our notation emphasizes that $\tbx$ is indeed the
Pryce coordinate (\ref{NWP}).},
\begin{subequations}
\begin{align}
\dot{\tbx}&=\hbp-e\hbar\Big(
\mu\,\frac{\hbp\times\bE}{|\bp|}
+
\partial_\bp(\mu\,\hbp\cdot\bB)
+
\mu\,\frac{\hbp\times(\hbp\times\bB)}{|\bp|}
\Big),
\label{motioneqnsr}
\\[6pt]
\dot{\bp}&=e\bE+{e}\hbp\times\bB
\label{motioneqnsp}
\\
&\quad+e\hbar\Big(-e\mu\frac{(\hbp\times\bE)\times\bB}{|\bp|}
+
\partial_{\tbx}(\mu\hbp\cdot\bB)
-e\partial_\bp(\mu\,\hbp\cdot\bB)\times \bB
-{e\mu}
\frac{(\hbp\cdot\bB)(\hbp\times\bB\big)}
{|\bp|}\Big),
\nn
\\[6pt]
\dot{\bS}&=
{e}(\frac{\bS}{2|\bp|})\times(\bE\times\hbp)
+
{e}\,(\frac{\bS}{2|\bp|})\times\bB,
\label{motioneqnsS}
\end{align} 
\label{motioneqns}
\end{subequations}
where the vector $\bS$  represents the spin and we introduced
$ 
\mu= \hbp\cdot\displaystyle\frac{\bS}{2|\bp|}\,. 
$ 
 Spin has now its own dynamics, (\ref{motioneqnsS}); 
 its feedback  to spatial motion is hidden in $\mu$.
 
These equations are similar to however different from  ours, (\ref{Prycevelocity}) - (\ref{ourpeq}) - (\ref{ourseq}).


Let us now turn to \emph{helicity}, our principal object of interest here. In \cite{BEPL}, it is defined as  the projection of the spin vector onto the momentum,
\beq
\lambda =\frac12\, \bS\cdot\hbp\,.
\label{lambdaheli}
\eeq
In the purely electric case the equations (\ref{motioneqnsr}) rewritten in  terms of $\lambda$
\cite{BEPL} look like those  in the non-commutative dynamics, (eqn. \# (1)-(2) of \cite{DHHMS}).  Consistency of the theory, namely the \emph{the Jacobi identity}  requires \cite{BMPRD}
\beq
{\rm div}_{\bp}\left(\lambda\frac{\bf p}{|\bf p|^3}\right)=0,
\label{Jacobidiv}
\eeq
which only allows for a constant $\lambda$, which it is not by (\ref{lambdaheli}).
Using (\ref{motioneqnsp}) and (\ref{motioneqnsS}) we found, moreover, that at order zero  in $\hbar$,
\beq
\frac{d\lambda}{dt}=-
\frac{e}{2} \left[\big(\hbp\times(\hbp\times\frac{\bS}{2|\bp|})\big)\cdot\bE+(\frac{\bS}{2|\bp|}\times\bB)\cdot\hbp \right],
\label{Bhelloss}
\eeq 
so that (\ref{lambdaheli}) is \emph{not conserved in general}. Further calculation (not reproduced here) indicates that $\lambda$ is not conserved let alone at the order of $\hbar$ \footnote{One could be tempted to define the helicity instead as
 $\mu=\lambda/|\bp|$. A simple calculation yields however that nor ${d\mu}/{dt}
$ does vanish at the order of $\hbar^0$.}. 

In conclusion, the model proposed in \cite{BEPL} exhibits the same type of helicity non-conservation as ours.
Let us also remark that the eqns.  
(\ref{motioneqns}) were obtained by  the approximation $E_p\approx |\bp|$. However, the
Hamiltonian, \# (12) of \cite{BEPL}, is
\beq
\label{BHam}
H_B=|\bp|
-
{e}(\frac{\hbar\bS}{2|\bp|})\cdot(\bE\times\hbp)
-e\,(\frac{\hbar\bS}{2|\bp|})\cdot\bB.
\eeq
Therefore $E_p$ should involve further, spin-field coupling terms as in (\ref{SFcE}), cf. Fig. \ref{dispersion}, yielding corrections to the equations of motion
\footnote{Note that (\ref{BHam}) is the weak-field approximation of \emph{our}
dispersion relation (\ref{SFcE}) \emph{with $g=1$}.}. 
\goodbreak

\section{Gravitational coupling}\label{gravcoupl} 

So far we considered a particle in flat space. As a further illustration, let us emphasize that, in the case of (minimal) gravitational coupling, the {helicity} is
still given by (\ref{Whel}).  
 Let us show how this comes about in general relativity.

The motion of a spinning particle in a curved background is given by the Papapetrou equations \cite{Papa51},
\begin{subequations}
\begin{align}
\label{dotP}
\dot{P}_\sigma&=-\frac{1}{2}\cR_{\mu\nu\rho\sigma}S^{\mu\nu}\dot{R}^\rho,
\\[2pt]
\label{dotS}
\dot{S}^{\mu\nu}&=P^\mu\dot{R}^\nu-P^\nu\dot{R}^\mu
\end{align}
\label{Papapetrou}
\end{subequations}
where the dot denotes the covariant derivative along the worldline and~$\cR=(\cR_{\mu\nu\rho\sigma})$ is the Riemann tensor. The non-deterministic character of the   equations (\ref{Papapetrou}) can be cured by imposing the standard relationship $S^{\mu\nu}P_\nu=0$, whose consistency with the Papapetrou equations enable us to posit the full set of constraints, 
\beq
S_{\mu\nu}P^{\nu}= 0,
\qquad
\half{S}_{\mu\nu}S^{\mu\nu}=s^2,
\qquad
P_{\mu}P^{\mu}=0,
\label{Gravconstraints}
\eeq
cf. (\ref{constraints}). Those imply
\begin{equation}
\dot{R}^\mu=P^\mu+\frac{2}{\Delta}S^{\mu\nu}\cR_{\alpha\beta\nu\rho}S^{\alpha\beta}P^\rho
\qquad\hbox{with}\qquad
\Delta=\cR_{\mu\nu\rho\sigma}S^{\mu\nu}S^{\rho\sigma}
\label{PapdotR}
\end{equation}
for some conveniently chosen parametrization  \cite{Sour74}. 
   Notice here the 
\textit{anomalous} spin--curvature - driven $4$-velocity, which is well-defined wherever $\Delta\neq0$.
The equations (\ref{Papapetrou}) and~(\ref{PapdotR}) can be obtained (using (\ref{Gravconstraints})) by the null foliation the two-form \cite{Sour74}
\begin{equation}
\sigma
=
-d^{\nabla}\!P_\mu\wedge{}dR^\mu
-\frac{1}{2s^2}\,d^{\nabla}\!{S}^\mu_\lambda\wedge{S}^\lambda_\rho\,d^\nabla\!{S}^\rho_{\;\mu}
-
\frac{1}{4}\cR_{\mu\nu\rho\sigma}S^{\mu\nu}dR^{\rho}{\wedge}\,dR^{\sigma}
\label{gravsigma}
\end{equation} 
generalizing (\ref{gsigma}) by the prescription of minimal gravitational coupling, $d\to{}d^{\nabla}$, where $d^{\nabla}$ denotes the exterior covariant derivative.

Using our  constraints (\ref{Gravconstraints}) we find 
 that $W$ and $P$ are again proportional, $W=\chi P$, 
for some function $\chi\neq0$ of the worldline. We thus have, on the one hand, $\star(S)W=\star(S)^2P=\left[S^2-\frac{1}{2}\Tr(S^2)\right]P=s^2P$ since $SP=0$ and $\Tr(S^2)=-2s^2$. On the other hand, we find $\star(S)W=\chi{}W=\chi^2P$, which finally implies 
\begin{equation}
{W}=\chi\,{P}
\qquad\hbox{with}\qquad
\chi=\pm|s|\,.
\label{chi=s}
\end{equation}
We thus confirm that
$\chi$, interpreted as the helicity of the minimally coupled massless particle, is indeed a constant in general relativity, as it is in special relativity.  

We mention for completeness that using (\ref{PapdotR}) Eq. (\ref{dotP})  simplifies  as \cite{Sat76}
\footnote{For any $F,S\in\ort(3,1)$, we have the general formula  $FSF=-\frac{1}{2}(S.F)\,F+\pf(F)\star(S)$, see \cite{Duval3these,Sour74}.} 
 
\begin{equation}
\dot{P}_\sigma=
-\frac{\pf(\cR(S))}{\Delta}\; W_\sigma\,,
\qquad
\label{dotPbis}
\end{equation}
where the Pfaffian of $\cR(S)_{\mu\nu}=\cR_{\alpha\beta\mu\nu}S^{\alpha\beta}$
is $\pf(\cR(S)))=
-\frac{1}{8}\sqrt{-g}\,\epsilon_{\alpha\beta\rho\sigma}\cR_{\gamma\delta}^{\;\;\;\;\alpha\beta}S^{\gamma\delta}
\cR_{\tau\omega}^{\;\;\;\;\rho\sigma}S^{\tau\omega}$. 

The \textit{monolocality} constraint $S_{\mu\nu}P^{\nu}=0$ in (\ref{Gravconstraints}) actually implies that $P_{\mu}P^\mu=\const$. No non-minimal gravitational coupling via the mass is therefore possible. One can furthermore couple the particle to an additional external electromagnetic field (Mathisson-Papapetrou equations \cite{Mathisson}) using the same constraint, and the vanishing of the electric dipole moment. This would lead to a mass formula $P_{\mu}P^\mu=f(eS.F)$ with $f$ an otherwise arbitrary function \cite{Duval3these,Sour74}, implying helicity non-conservation through spin-field interaction, as in (\ref{constraints}).

\section{Conclusion}

This Letter is devoted to the study of \emph{classical helicity}.
 We found, within the S-model \cite{DHchiral,SSD}, that   $\chio$ in (\ref{oldhelicity}) and its generalization ${\chi}$ in (\ref{Whel}) are the same constants of the motion when the  particle is free or when it is coupled \emph{minimally} to an electromagnetic or to a gravitational field. But none of them is conserved in general, when  coupled \emph{non minimally}.
 
For non-vanishing effective mass, neither 
 of our expressions is boost-invariant, cf. (\ref{helPoincare}), consistently with what is known for massive particles \cite{Peskin}.
 
This anomalous behavior, illustrated by motion for $g=2$ in a pure electric field, confirms that the spin-field coupling (\ref{effmass}) converts the massless system to an effectively massive one. 


Further aspects including field theory and electric-magnetic duality are in progress and will be reported elsewhere.

\begin{acknowledgments} 
PH would like to thank K. Bliokh for discussions at the SIS'15 meeting in Yerevan (Armenia). ME and PH are grateful to the IMP of the CAS for hospitality in Lanzhou. 
 This work was supported by the Major State Basic Research Development Program in China (No.
2015CB856903) and the National Natural Science Foundation of China (Grant No. 11035006 and 11175215). 
\end{acknowledgments}
\goodbreak


\end{document}